\documentclass[10pt]{article}
\usepackage{times}
\usepackage{epsf}
\newcommand{\figcap}[1]{
\it
\caption{#1}
  }
\newcommand{\dd}{\mbox{d}}
\newcommand{\DD}{\mbox{D}}
 
\newcommand{\tfrac}[2]{{\textstyle\frac{#1}{#2}}}

\begin{document}
\begin{titlepage}
\title{String Picture \\ of \\Bose-Einstein Condensation} \author{S.\
Bund and Adriaan M.\ J.\ Schakel\thanks{E-mail:
schakel@physik.fu-berlin.de} \\ Institut f\"ur Theoretische Physik \\
Freie Universit\"at Berlin \\ Arnimallee 14, 14195 Berlin }
\date{\today}
\maketitle
\begin{abstract}
A nonrelativistic Bose gas is represented as a grand-canonical ensemble
of fluctuating closed spacetime strings of arbitrary shape and length.
The loops are characterized by their string tension and the number of
times they wind around the imaginary time axis.  At the temperature
where Bose-Einstein condensation sets in, the string tension, being
determined by the chemical potential, vanishes, the system becomes
critical, and the strings proliferate.  A comparison with Feynman's
description in terms of rings of cyclicly permuted bosons shows that the
winding number of a loop corresponds to the number of particles
contained in a ring.
\end{abstract}
\end{titlepage}

The first experimental realization of Bose-Einstein condensation (BEC) in a
cloud of weakly interacting $^{87}$Rb atoms trapped in a magnetic field by a
group in Boulder, Colorado \cite{Boulder}, and the second definite
observation of BEC in a system of Na atoms at MIT \cite{MITBEC}, turned this
field into one of the most active research areas in contemporary condensed
matter physics.  This recent burst of interest motivated us to consider once
more the fundamental properties of BEC.

At present there exist at least three different descriptions of this
phenomenon.  In the first, and intuitively probably the simplest one, BEC is
understood as the condensation of a finite fraction of the total number of
particles present in the zero-momentum ground state \cite{Huang}.  The
second description, due to Bogoliubov \cite{Bogoliubov}, uses the methods of
quantum field theory \cite{Brown}.  The formation of a Bose-Einstein
condensate is here signaled by a nonzero expectation value of the field
describing the particles in the system.  The third description, due to
Feynman \cite{lambda}, is based on path integrals \cite{Feynman48}.  BEC is
now understood as the occurrence of large rings containing many bosons which
are cyclicly permuted in imaginary time \cite{statmech,Stone}.  Using a
mapping onto a certain problem of classical statistical mechanics
\cite{ChanWol} to numerically evaluate the path integral, Ceperley and
Pollock \cite{CePo} showed that this picture also provides a powerful
computational tool with which even a strongly interacting system like
superfluid $^4$He can be accurately described (for a review, see Ref.\
\cite{Ceperley}).  The classical system is a chain of beads connected to
springs.

In this Letter we discuss an alternative, string picture of BEC.  This
description is, like Feynman's, based on the spacetime approach to
quantum mechanics, i.e., on path integrals.  The specific path-integral
representation we arrive at was first put forward by Montroll and Ward
\cite{MoWa}.  Physically, it describes a grand-canonical ensemble of
fluctuating closed spacetime strings of arbitrary shape and length.  We
study the critical behavior of this loop gas close to the temperature
$T_{\rm c}$ where Bose-Einstein condensation sets in, and show that the
loops proliferate below $T_{\rm c}$.  That is to say, the loop gas
undergoes a Hagedorn transition at $T_{\rm c}$.  

The critical behavior we found here is similar to that of vortex loops
in superfluids \cite{ABH} and in type-II superconductors \cite{NgSu}.
(For a detailed account of the so-called dual approach to superfluidity
and superconductivity in which the vortex loops are the elementary
excitations, the reader is referred to the textbook by Kleinert
\cite{GFCM}.)

Our starting point is the Lagrangian of a nonrelativistic free scalar
field theory in Euclidean spacetime $x=(\tau,{\bf x})$, with $\tau$ the
imaginary time \cite{Brown}:
\begin{equation}
{\cal L} = \phi^*\left(\hbar\partial_\tau - \mu -
\frac{\hbar^2}{2m}\nabla^2\right)\phi .
\end{equation}
Here, $\phi$ is the field describing the scalar particles, $\partial_\tau =
\partial/\partial \tau$, $\mu$ is the chemical potential which accounts for
a finite particle number density, and $-\hbar^2\nabla^2/2m$, with $m$ the
mass of the particles, is the kinetic energy operator.  Depending on whether
the particles are bosons or fermions, the field $\phi$ is an ordinary
commuting or an anticommuting Grassman field.

We adopt the imaginary-time approach to thermal field theory, where the time
variable $\tau$ is taken to be of finite extent $0 \leq \tau \leq \hbar
\beta$, with $\beta$ the inverse temperature $\beta = 1/k_{\rm B}
T$ ($k_{\rm B}$ is Bolzmann's constant) \cite{Rivers}.  As a result, the
energy variable conjugate to $\tau$ takes on discrete values.  Depending on
whether $\phi$ describes bosons or fermions, the field is periodic or
antiperiodic in $\tau$, $\phi(\hbar \beta,{\bf x}) = \pm \phi(0,{\bf x})$,
and can be expanded in a Fourier series as:
\begin{equation} 
\phi (x) = \frac{1}{\hbar \beta} \sum_{n=-\infty}^\infty \int \frac{\dd^3
k}{(2\pi)^3} {\rm e}^{- i k \cdot x} \phi (k),
\end{equation}  
where $k = (\omega_n,{\bf k})$, $k \cdot x = \omega_n \tau - {\bf k} \cdot
{\bf x}$, and $\omega_n$ are the discrete Matsubara frequencies,
\begin{equation} 
\omega_n = \left\{ \begin{array}{ll} 2n \pi/ \hbar\beta & \mbox{for \,
                 bosons} \\ (2n + 1) \pi/ \hbar \beta & \mbox{for \,
                 fermions}, \end{array} \right. 
\end{equation} 
with $n$ an integer.

At finite temperature and density, the thermodynamic properties of the system
are specified by the partition function $Z$.  Expressed as a functional
integral over the field $\phi$ it reads \cite{Brown}:
\begin{equation} 
Z = \int \DD \phi^* \DD \phi \, \exp\left(-\frac{1}{\hbar} \int_0^{\hbar
\beta} \dd \tau \int \dd^3 x \, {\cal L}\right).
\end{equation} 
Since this functional integral is Gaussian, it is easily evaluated to yield
\begin{eqnarray}   \label{Gauss}
\ln (Z) &=& \ln \left\{ {\rm Det}^{-\eta}\left[\left(-i\hbar\omega_n - \mu+
\hbar^2 {\bf k}^2/2m\right)/\hbar\right] \right\} \nonumber \\ &=& - \eta
{\rm Tr} \ln \left[\left(-i\hbar\omega_n -\mu + \hbar^2 {\bf
k}^2/2m\right)/\hbar\right],
\end{eqnarray} 
where we used the identity Det(A) = exp[Tr ln(A)].  In Eq.\ (\ref{Gauss}),
$\eta = \pm 1$ for bosons and fermions, respectively, and the trace Tr
stands for the integral over Euclidean spacetime $x$ as well as the
summation over Matsubara frequencies $\omega_n$ and the integral over
momentum ${\bf k}$.

We wish to derive an alternative representation of the system and express
$Z$ not as a {\it functional} integral over the field $\phi$, but as a {\it
path} integral over spacetime loops.  In the string picture that emerges,
the system is characterized by the string tension of the loops and the
number of times they wind around the imaginary time axis.

Before considering the partition function, it is expedient to first study
the propagator $G(x_1,x_2) = G(x_1 - x_2)$ of the theory,
\begin{equation}  \label{propagator}
G(x) = \frac{1}{\hbar\beta}\sum_{n=-\infty}^\infty \int \frac{\dd^3 k}{(2
\pi)^3} \frac{\hbar} {-i\hbar\omega_n - \mu + \hbar^2 {\bf k}^2/2m} \, {\rm
e}^{-i k \cdot x},
\end{equation} 
and write it as a path integral over spacetime trajectories.  Employing
Schwinger's proper-time method \cite{proptime}, which is based on Euler's
form
\begin{equation} \label{Euler}
\frac{1}{a^z} = \frac{1}{\Gamma(z)} \int_0^\infty \frac{\dd s}{s} \,
s^z \, {\rm e}^{- s a},
\end{equation} 
with $\Gamma(z)$ the Gamma function, we write the right-hand side of
Eq.\ (\ref{propagator}) as an integral over proper time $s$:
\begin{equation} 
G(x) = \int_0^\infty \dd s \, {\rm e}^{\mu s/\hbar} \,
\frac{1}{\hbar\beta}\sum_n {\rm e}^{-i\omega_n(\tau -s)} \int \frac{\dd^3
k}{(2 \pi)^3} \, {\rm e}^{i {\bf k} \cdot {\bf x} - \hbar {\bf k}^2 s/2m}.
\end{equation} 
We next use Poisson's summation formula,
\begin{equation} \label{Poisson}
\sum_{n=-\infty}^\infty {\rm e}^{2\pi i n\alpha} =
\sum_{w=-\infty}^\infty \delta(\alpha-w)
\end{equation}
to replace the summation over the Matsubara frequencies by another
one
\begin{equation}  \label{tau-s}
\frac{1}{\hbar\beta} \sum_{n=-\infty}^\infty {\rm e}^{-i\omega_n(\tau - s)}
= \sum_{w=-\infty}^\infty \eta^w\delta[s-(\tau+w\hbar\beta)].
\end{equation} 
The physical meaning of the new summation index will become clear shortly.
The delta function in Eq.\ (\ref{tau-s}) links Schwinger's proper time $s$
to the Euclidean time $\tau$.  The propagator now becomes:
\begin{equation}   \label{prop}
G(x) = \sum_{w=0}^\infty\eta^w
\exp\left[\frac{\mu}{\hbar}(\tau+w\hbar\beta)\right] \left[\frac{m}{2 \pi
\hbar(\tau+w\hbar\beta)}\right]^{3/2} \exp\left[-\frac{m {\bf
x}^2}{2\hbar(\tau+w\hbar\beta)}\right].
\end{equation} 
According to the path-integral approach to quantum mechanics
\cite{Feynman48}, this can be written as a sum over all possible spacetime
trajectories running from ${\bf x}(0)=0$ to ${\bf x}(\tau+w\hbar\beta)={\bf
x}$:
\begin{equation}   \label{path}
G(x) = \sum_{w=0}^\infty \eta^w G_w(x),
\end{equation} 
with
\begin{equation} \label{alias}
G_w(x) = \int_{{\bf x}(0)=0}^{{\bf x}(\tau+w\hbar\beta)={\bf x}} \DD{\bf
x}(\tau')\, \exp\left(-\frac{1}{\hbar}\int_0^{\tau+w\hbar\beta} \dd \tau'\,
L \right).
\end{equation} 
Here, $L$ is the (Euclidean) Lagrangian of a nonrelativistic point particle
with mass $m$ in a constant potential $-\mu$:
\begin{equation} 
L = \tfrac{1}{2}m \dot{{\bf x}}^2(\tau) - \mu,
\end{equation} 
where $\dot{\bf x}(\tau) = \partial {\bf x}/ \partial \tau$.  Remembering
the (anti)periodicity in $\tau$ with period $\hbar \beta$, we see that the
summation index $w$, which was introduced in Eq.\ (\ref{tau-s}) to replace
the summation over the Matsubara frequencies, denotes the number of times a
given trajectory winds around the (Euclidean) time axis.  The sum over the
winding number $w$ in Eq.\ (\ref{path}) then implies that the trajectories,
which started at ${\bf x}(0) = 0$, can wrap arbitrarily many times around the
time axis before reaching their end point ${\bf x}(\tau+w\hbar\beta)={\bf
x}$.

We next turn to the partition function.  The two main differences with the
propagator are that, first, instead of a factor 1/A in Eq.\
(\ref{propagator}) we have a ln(A) in Eq.\ (\ref{Gauss}).  And second, the
expression for the partition function contains, in contrast to that for the
propagator, no $x$-dependence.  Taking these differences into account, we
obtain
\begin{equation} 
\ln (Z) = \eta V \int_0^\infty \frac{\dd s}{s} {\rm e}^{\mu s/\hbar}
\sum_{n=-\infty}^\infty {\rm e}^{i\omega_ns} \int \frac{\dd^3 k}{(2 \pi)^3}
\, {\rm e}^{-i \hbar {\bf k}^2 s/2m},
\end{equation} 
where we used Euler's form (\ref{Euler}) in the limit of small $z$,
\begin{equation} \label{Slog}
\ln(a) = - \int_0^\infty \frac{\dd s}{s} \, {\rm e}^{- s a} ,
\end{equation} 
ignoring an irrelevant additive constant.  The integration over spacetime
produced a factor $\hbar \beta V$, with $V$ the volume of the system.
Following the steps leading to the explicit expression (\ref{prop}) for the
propagator, we arrive at the familiar fugacity $\exp(\beta \mu)$ series
\cite{Huang}
\begin{equation} \label{explicit}
\ln(Z) = \frac{V}{\lambda^3} \sum_{w=1}^\infty \eta^{w+1} \frac{{\rm
e}^{w\beta\mu}}{w^{5/2}},
\end{equation} 
with 
\begin{equation} 
\lambda  = \hbar \sqrt{2 \pi \beta/m} 
\end{equation} 
the de Broglie thermal wavelength.  In Eq.\ (\ref{explicit}), we omitted the
singular zero-tempera\-ture contribution corresponding to zero winding number
$w=0$.

The corresponding path-integral representation of $\ln(Z)$ is obtained from
Eq.\ (\ref{explicit}) in a similar way as the path-integral representation
(\ref{path}) of the propagator was obtained from Eq.\ (\ref{prop}).  We
find:
\begin{equation}  \label{lnZ}
\ln(Z) = V \sum_{w=1}^\infty \eta^{w+1} G_w(0)/w ,
\end{equation} 
where $G_w$ was expressed as a path integral in Eq.\ (\ref{alias}).  The
extra factor $1/w$ stems from the factor $1/s$ in the Schwinger
representation (\ref{Slog}) of the logarithm, while the zero argument in
$G_w(0)$ reflects the absence of any spacetime dependence in the partition
function.  Explicitly,
\begin{equation} \label{closed}
G_w(0) = \int_{{\bf x}(0)=0}^{{\bf x}(w\hbar\beta)=0} \DD{\bf x}(\tau')\,
\exp\left(-\frac{1}{\hbar}\int_0^{w\hbar\beta} \dd \tau'\, L \right),
\end{equation} 
implying that this path integral is a sum over all possible {\it closed}
spacetime trajectories, or loops, that return to their starting point
after a time $w\hbar\beta$.  Since one can start traversing a closed
trajectory anywhere along the loop, the extra factor $1/w$ (or, more
precisely, $1/w\hbar\beta$) in Eq.\ (\ref{lnZ}) can be intuitively
understood as arising to prevent overcounting.  The arbitrariness of the
space coordinate of the starting point yields the volume factor $V$.

In summary, we expressed the partition function as a path integral over
spacetime loops.  Each loop in the loop gas can be of arbitrary shape and
length, and is characterized by a winding number $w$, showing how often it
wraps around the time axis.  The sum over $w$ in Eq.\ (\ref{lnZ}) implies
that, in principle, this can be arbitrary many times.

\begin{figure}
\begin{center}
\epsfxsize=8.cm \mbox{\epsfbox{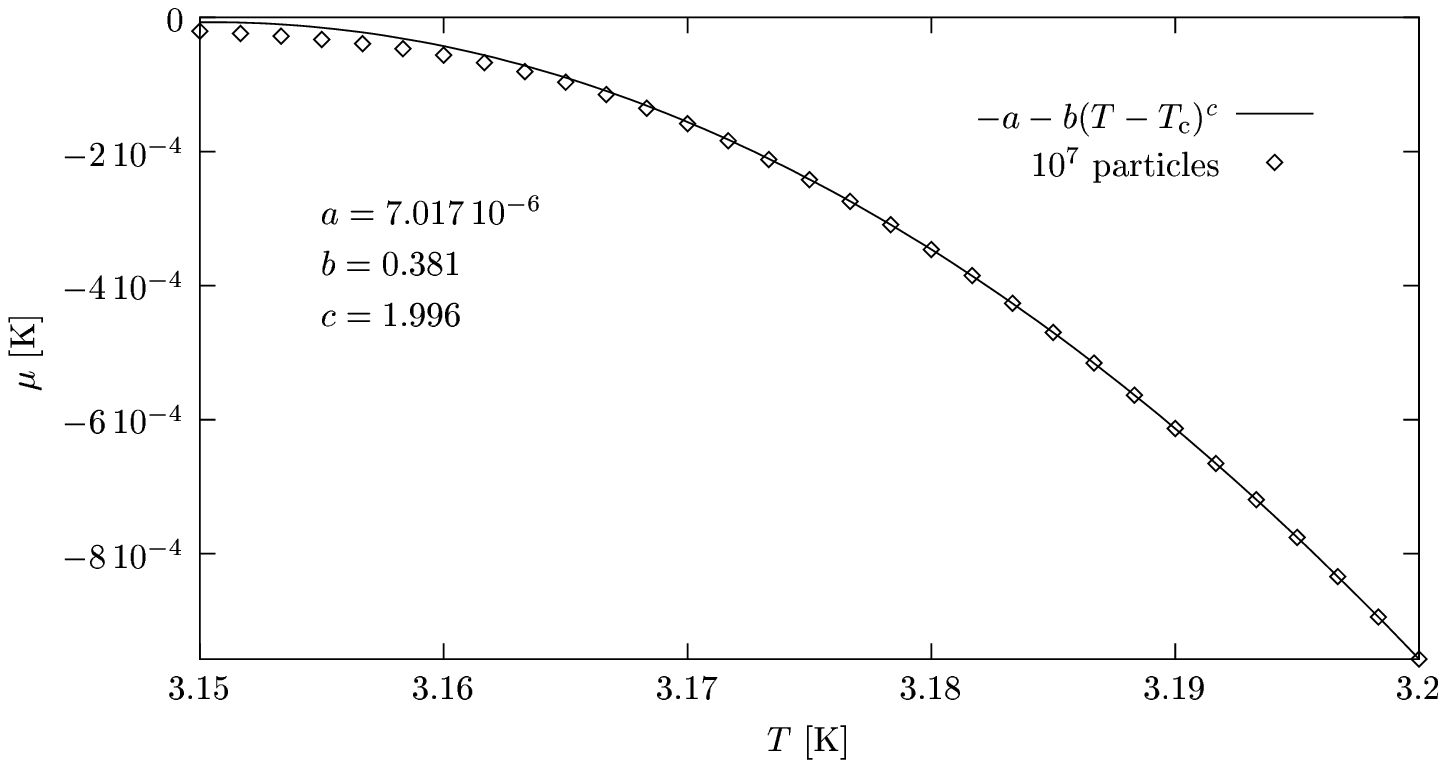}}
\end{center}
\figcap{Fit of the chemical potential. \label{fig:fit}}
\end{figure}

With this insight, we return to the explicit expression (\ref{explicit}) of
the partition function, restricting ourselves to bosons.  To be able to
carry out numerical studies, we put the system in a box with sides of length
$D$, so that $V = D^3$.  As a result, the integration over momentum is
replaced by a summation over the discrete variable ${\bf k}_{{\bf n}} =
(2\pi/D) {\bf n}$, where ${\bf n}$ denotes the integers $(n_x, n_y, n_z)$.
This leads in turn to the following change in the propagator (\ref{prop}):
\begin{equation} 
\exp\left[-\frac{m {\bf x}^2}{2\hbar(\tau+w\hbar\beta)}\right]
\rightarrow \sum_{{\bf n}} \exp\left[-\frac{m}{2\hbar
(\tau+w\hbar\beta)}({\bf x}-{\bf n} D)^2\right].
\end{equation} 
Instead of Eq.\ (\ref{explicit}), we now obtain as explicit form for the
partition function
\begin{equation} \label{explicit'}
\ln(Z) = \frac{D^3}{\lambda^3} {\sum_w}'\frac{{\rm e}^{w\beta\mu}}{w^{5/2}}
W^3\left(\frac{\pi}{w} \frac{D^2}{\lambda^2}\right),
\end{equation} 
where the prime on the sum will be explained below and
\begin{equation}
W(x) = \sum_{n=-\infty}^\infty {\rm e}^{-x n^2}.
\end{equation} 
The average particle number density $n = \langle N \rangle/V$ in the box is
obtained by differentiating $\ln(Z)$ with respect to the chemical potential:
\begin{equation} \label{N}
n = \frac{1}{\beta}\frac{\partial \ln(Z)}{\partial\mu}.
\end{equation} 
In the following we shall assume this density to be fixed.  As numerical
value we take $n = 22.22$ nm$^{-3}$, which is the density of liquid $^4$He
at saturated vapor pressure.  When varying the size of the system, we always
adjust the particle number $\langle N \rangle = N$ in such a way that the
density remains constant.  For a given volume, Eq.\ (\ref{N}) then
determines the chemical potential as function of $N$ and $\beta$.  Figure
\ref{fig:fit} shows the temperature dependence of $\mu$ at fixed particle
number $N=10^7$.  We observe that it exhibits critical behavior, tending to
zero as
\begin{equation}  \label{fit}
\mu(T) \sim -(T-T_{\rm c})^c,
\end{equation} 
with the critical exponent $c=1.996$.  Here, $T_{\rm c}$ is the condensation
temperature of a free Bose gas,
\begin{equation} 
T_{\rm c} = \frac{2 \pi \hbar^2}{k_{\rm B} m} \left[
\frac{n}{\zeta(3/2)}\right]^{2/3},
\end{equation} 
where $\zeta$ is the Riemann zeta-function, with $\zeta(3/2) = 2.61238
\cdots$.  Numerically, taking the $^4$He value also for the mass parameter,
$m = 6.695 \; 10^{-27}$ kg, one finds $T_{\rm c} = 3.145$ K.

The exponent $c$ is related to the correlation length exponent $\nu$ and the
dynamic exponent $z$ via
\begin{equation} 
c = 2 \nu z.
\end{equation} 
The dynamic exponent appears here because the strings are embedded in
spacetime and parameterized by Euclidean time. (For strings embedded in
space, we would have $c = 2 \nu$ instead.)  The value $c \approx 2$ we found
numerically agrees with the expected Gaussian exponents $\nu =1/2$ and
$z=2$.

Since $\ln(Z)$ is written in Eq.\ (\ref{explicit'}) as a sum over the
winding number $w$, also $N$ can be expressed in this way:
\begin{equation}  \label{specify}
N = {\sum_w}' \langle N_w \rangle ,
\end{equation} 
where
\begin{equation} \label{Nw}
\langle N_w \rangle = \frac{D^3}{\lambda^3} \frac{{\rm
e}^{w\beta\mu}}{w^{3/2}} W^3\Bigl(\frac{D^2}{w\lambda^2}\Bigr)
\end{equation} 
denotes the average number of particles contained in loops with winding
number $w$.  To obtain a loop interpretation of the total particle number
$N$, note that $\langle N_w \rangle$ is simply related to the average number
$\langle L_w \rangle$ of loops with winding number $w$ via
\begin{equation}  \label{Lw}
\langle L_w \rangle = \frac{\langle N_w \rangle}{w},
\end{equation} 
so that 
\begin{equation} \label{looplength}
N = {\sum_w}' w \langle L_w \rangle.
\end{equation} 
This means that the particle number $N$ is given by the total loop
``length'' measured in units of the characteristic ``length'' scale $a
=\hbar \beta$.  We used quotation marks here because the loops are
embedded in spacetime and parameterized by Euclidean time, so that $a$ has the
dimension of time, not of length.  

Since we keep the number of particles in the box fixed, the summations in
Eqs.\ (\ref{specify}) and (\ref{looplength}) have the upper bound $N$; there
are no particles available to fill longer loops.  To indicate this we
decorated the Sigma's with a prime.

The partition function written in terms of loops reads
\begin{equation} 
\ln(Z) = {\sum_w}' \langle L_w \rangle.
\end{equation} 
It follows that each loop gives the same contribution to the pressure $P=
\ln (Z)/\beta V$, and also to the energy density
\begin{equation} 
e = - \left. \frac{\partial (\beta P)}{\partial \beta} \right|_{\beta \mu} =
\frac{3}{2} \frac{1}{\beta V} {\sum_w}' \langle L_w \rangle
\end{equation} 
of the system, independent of the loop length or, equivalently, the number
of particles contained in the loop.

\begin{figure}
\begin{center}
\epsfxsize=8.cm  \mbox{\epsfbox{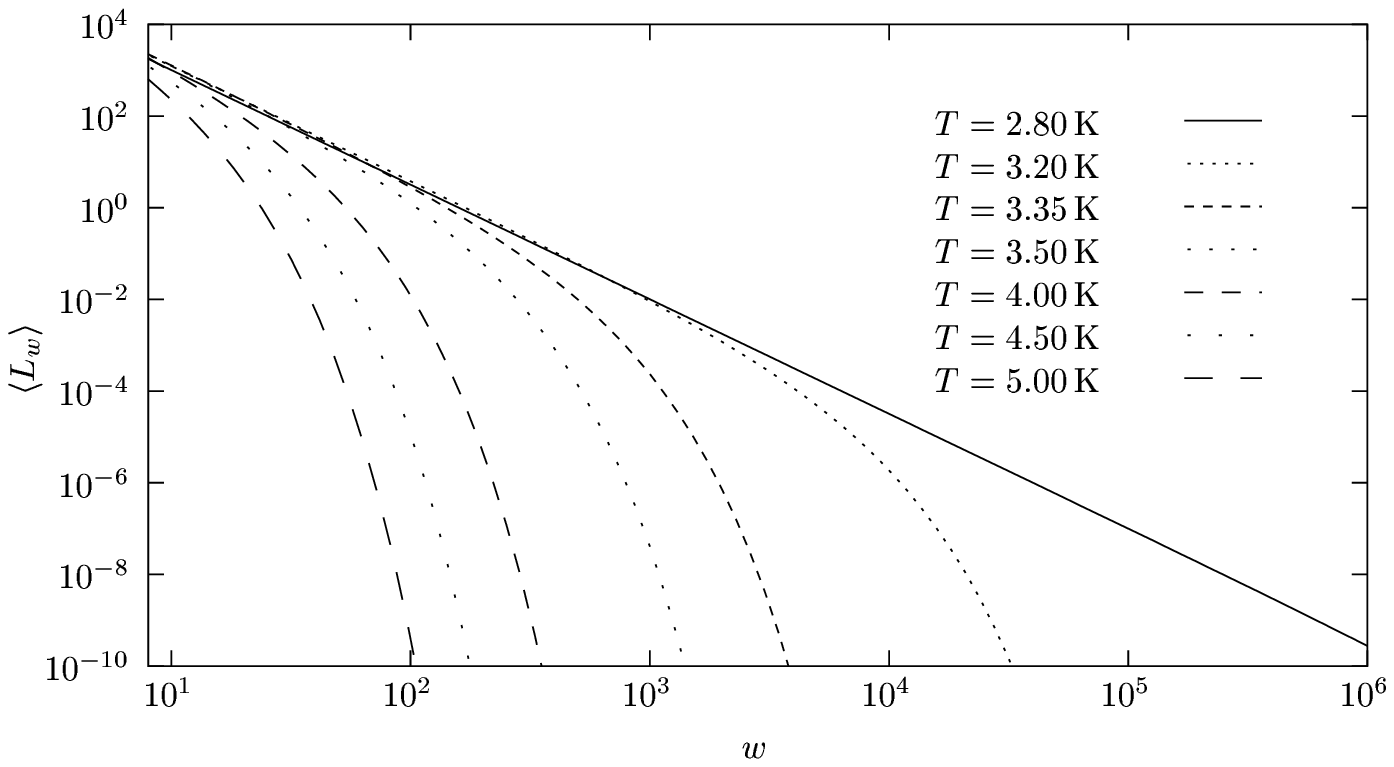}}
\end{center}
\figcap{Loop length distribution for various
temperatures. \label{fig:distr}}
\end{figure}

In the limit of large $D$, the loop length distribution (\ref{Lw}) with Eq.\
(\ref{Nw}) reduces to
\begin{equation} \label{loopdistribution}
\langle L_w \rangle = \left(\frac{m \beta}{2 \pi}\right)^{3/2}
\left(\frac{D}{a}\right)^3 \left(\frac{\ell}{a}\right)^{-5/2} {\rm e}^{-
\beta \sigma \ell},
\end{equation} 
where $\ell = w a$ is the ``length'' of the loop measured in units of
$a=\hbar \beta$, and $\sigma = -\mu/\hbar \beta$ is the string
``tension''.  Equation (\ref{loopdistribution}) is the typical form for
the loop length distribution of Brownian strings \cite{Copetal}.  Figure
\ref{fig:distr} shows this distribution for various temperatures,
assuming again liquid $^4$He values for the mass parameter $m = 6.695 \;
10^{-27}$ kg and the (fixed) particle number density $n = 22.22$
nm$^{-3}$.
For $T > T_{\rm c}$, the chemical potential is negative and the strings
have a positive tension.  In this temperature regime, we see an
exponential decay.  When the condensation temperature $T_{\rm c}$ is
approached from above, the chemical potential and thus the string
tension vanishes.  The decay is seen to change from exponential to
algebraic.  Because their tension vanishes, the strings proliferate.
(The temperature where this happens is known in this context as the
Hagedorn temperature.)  BEC, therefore, corresponds to the appearance of
long loops, wrapping arbitrarily many times around the time axis
\cite{Stone}.  

A similar proliferation of line-like objects we found here, occurs in
superfluids and type-II superconductors \cite{GFCM}.  The line-like
objects in these systems are vortices which are the elementary
excitations of the dual approach.  In particular, the string tensions
and the loop length distributions behave similarly \cite{ABH,NgSu}.
Only the numerical values of the critical exponents differ as a
superfluid and a type-II superconductor belong to the $XY$ universality
class, whereas a free Bose gas has Gaussian exponents.  Another
difference is that a dual transformation interchanges the low- and
high-temperature phases, so that the vortices proliferate in the
high-temperature phase.

\begin{figure}
\begin{center}
\epsfxsize=8.cm
\mbox{\epsfbox{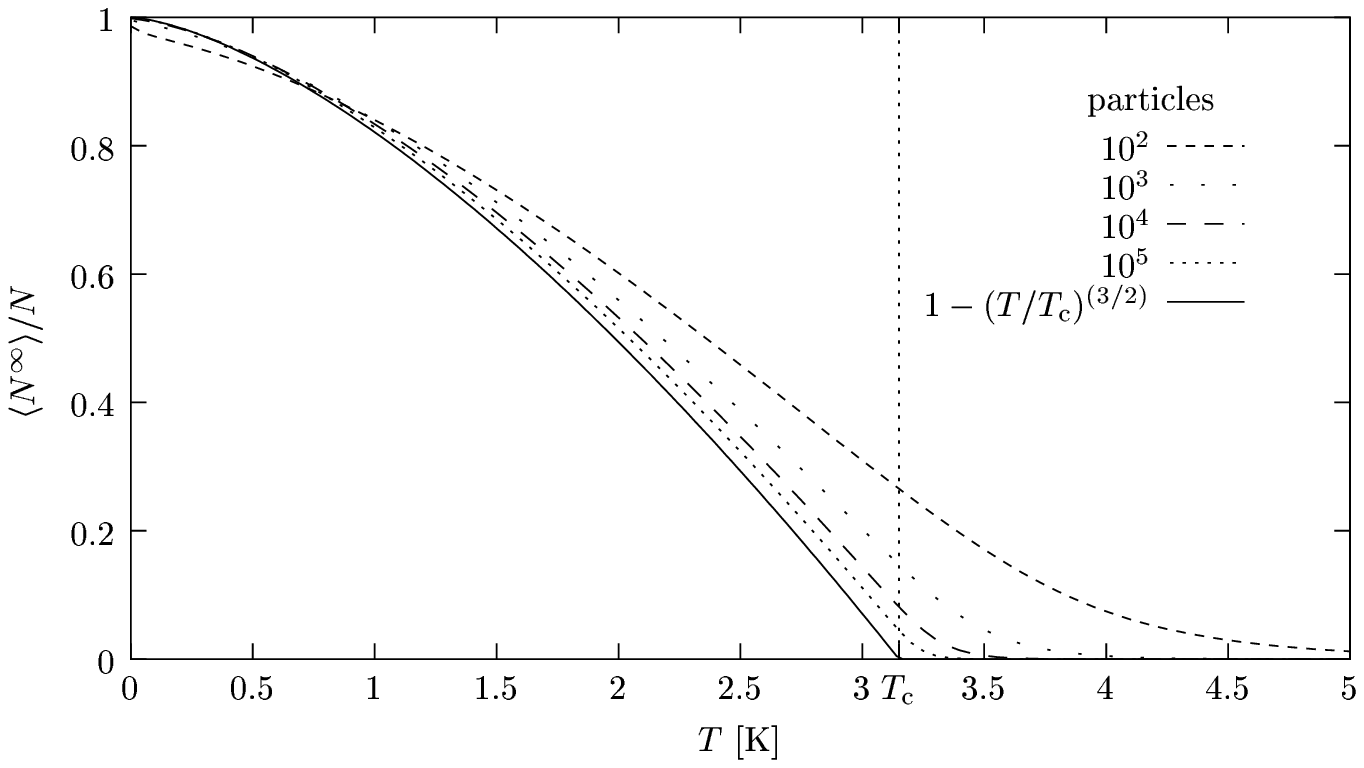}}
\end{center}
\figcap{Fraction of particles contained in long loops.  \label{fig:infi}}
\end{figure}

To study numerically the correspondence between BEC and the appearance of
long or, for an infinite system, infinite loops, we follow Ref.\ \cite{ABH}
and distinguish two types of loops: small ones with winding number $w <
\sqrt{N}$, and long ones with $w > \sqrt{N}$.  The choice $w=\sqrt{N}$ as
boundary is motivated by our observation that the average winding number
squared $\langle w^2 \rangle$, which gives the compressibility of the
system, makes a jump of the order of $N$ at the condensation temperature.
Figure \ref{fig:infi} shows the temperature-dependence of the average number
$\langle N^\infty\rangle$ of particles contained in long loops,
\begin{equation}
\langle N^\infty\rangle = \sum_{w=\sqrt{N}}^N \langle N_w \rangle,
\end{equation} 
as fraction of the total number of particles present.  Because of Eq.\
(\ref{looplength}), revealing that the particle number in the string picture
is nothing but the total loop length measured in units of $a=\hbar \beta$,
Fig.\ \ref{fig:infi} can be equivalently interpreted as showing the
temperature-dependence of the combined length of long loops as fraction of
the total loop length.
We see that long loops are exponentially suppressed in the normal phase,
implying that here only small loops containing just a few particles are
present.  The larger the system is, the more suppressed the long loops are.
Below the condensation temperature, long loops appear in the system.  We see
their number increasing with decreasing temperature.  At the absolute zero
of temperature, virtually all particles are contained in long loops.  With
increasing particle number $N$, the curves converge to the solid line given
by
\begin{equation} \label{Ninfty}
\frac{\langle N^\infty \rangle}{N}\Biggr|_{N\to\infty} = \left\{
\begin{array}{ll} 1-(T/T_{\rm c})^{3/2} & \mbox{for $T<
T_{\rm c}$} \\ 0 & \mbox{for $T>T_{\rm c}$}, \end{array} \right.
\end{equation} 
implying that also this loop length, like the string tension, exhibits
critical behavior at $T_{\rm c}$.  The right-hand side of Eq.\
(\ref{Ninfty}) is the well-know expression for the number of condensed
particles.  From this we conclude that the particles contained in long loops
are condensed in the zero-momentum ground state, so that BEC indeed
corresponds in the string picture to the appearance of long loops, i.e., to
the proliferation of strings.

\begin{figure}
\begin{center}
\epsfxsize=8.cm
\mbox{\epsfbox{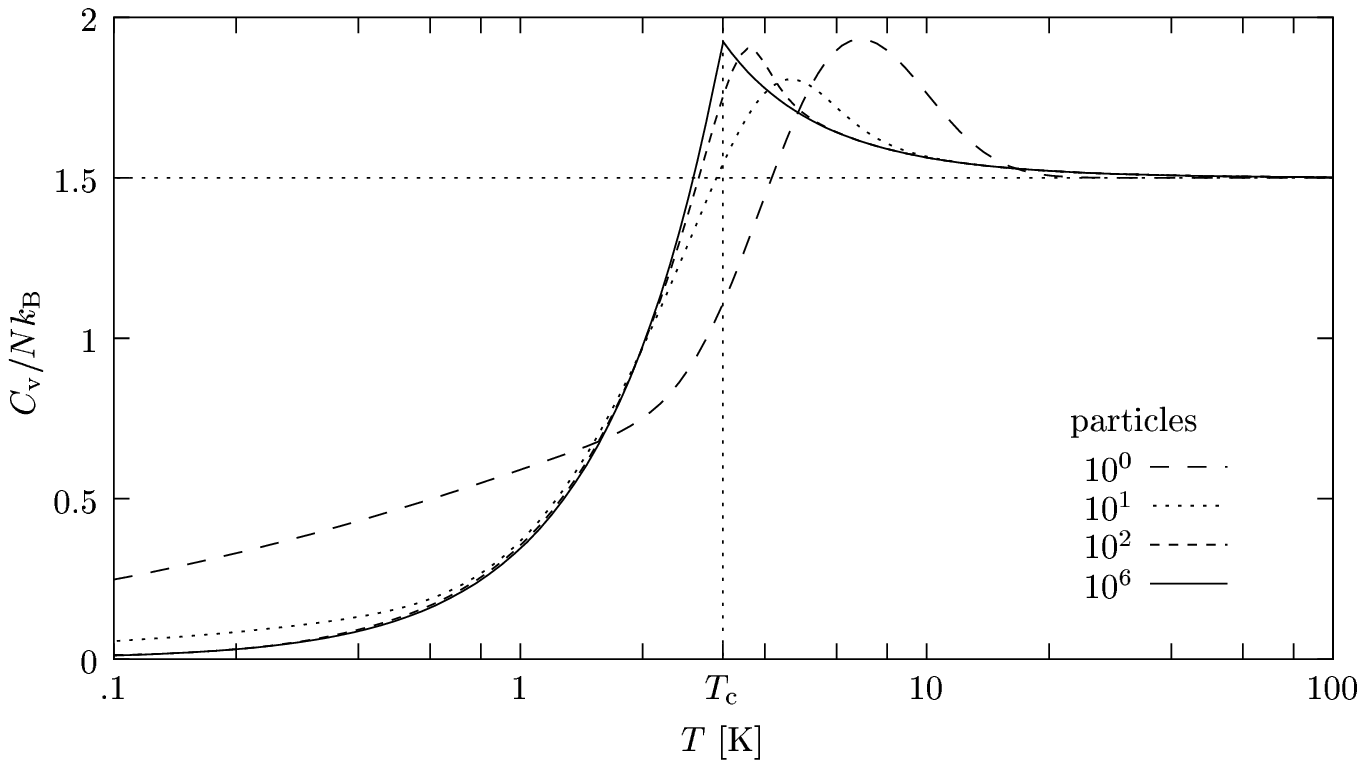}}
\end{center}
\figcap{Heat capacity for different numbers of particles. \label{fig:cv}}
\end{figure}

An interesting question is how many particles are needed for BEC.  We
investigate this point numerically by calculating the heat capacity for
various values for the particle number $N$ (see Fig. \ref{fig:cv}).
We observe that already for a particle number of the order 10, the behavior of
the heat capacity at low temperatures closely resembles that of a system
with many particles.  From this we conclude that about 10 particles suffice
for BEC to arise. 

\begin{figure}
\begin{center}
\epsfxsize=10.cm
\mbox{\epsfbox{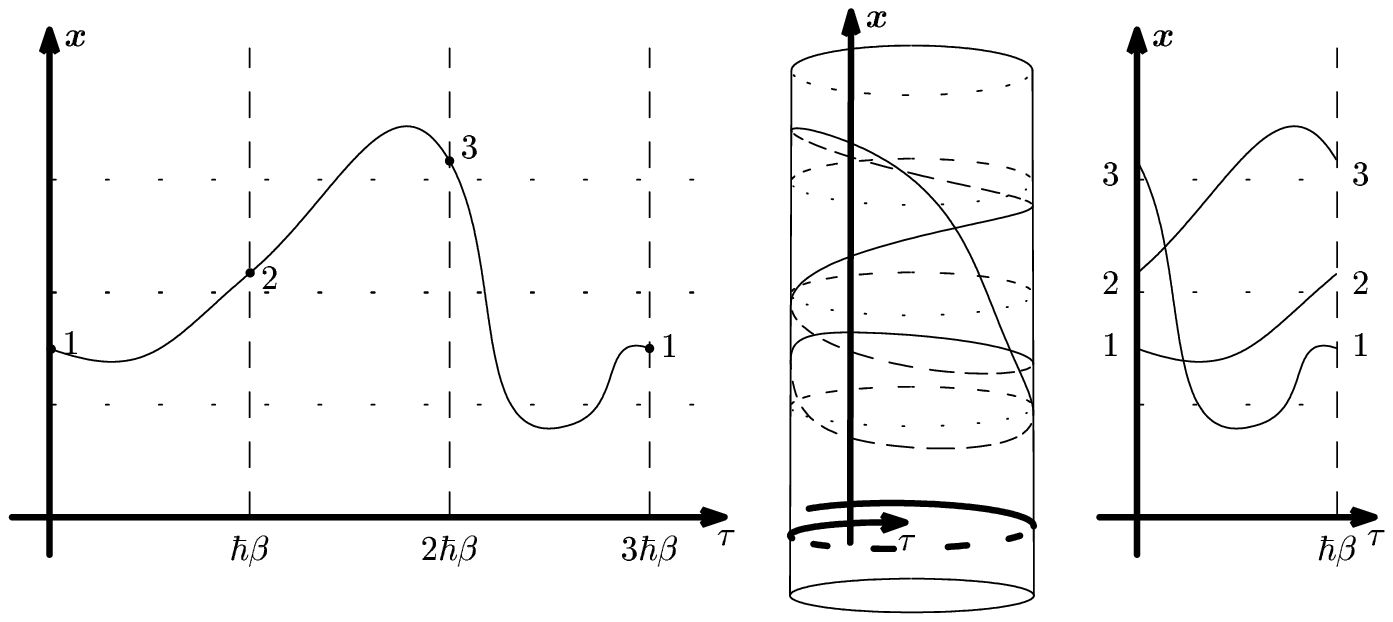}}
\end{center}
\figcap{Interpretation of a loop with winding number $w=3$ as the worldlines
of three cyclicly permuted particles. \label{fig:many}}
\end{figure}

In the final part of this Letter, we wish to establish contact with
Feynman's path-integral representation of the partition function of a
nonrelativistic system and its interpretation in terms of rings of
cyclicly permuted bosons \cite{statmech}.  To this end, we recast
Eq. (\ref{lnZ}) in the form:
\begin{equation} \label{Z}
Z = \sum_{l=1}^\infty \frac{V^l}{l!}  \sum_{w_1,\dots,w_l=1}^\infty
\left[\eta^{w_1+1} G_{w_1}(0)/w_1\right] \cdots
\left[\eta^{w_l+1}G_{w_l}(0)/w_l\right],
\end{equation} 
where the summation over $l$ arises after expanding the exponential function
in a Taylor series; $l$ physically denotes the number of loops.  We omitted
the constant term corresponding to $l=0$.  It is helpful to introduce the
constraint $\sum_{j=1}^l w_j=N$ together with an extra summation over $N$,
so that
\begin{equation}  \label{extra}
\sum_{w_1,\dots,w_l=1}^\infty \rightarrow \sum_{N=1}^\infty\;
\sum_{\begin{array}{c} \scriptstyle w_1,\dots,w_l=1 \\ \scriptstyle
\sum_j w_j=N \end{array}}^\infty
\end{equation} 
in Eq.\ (\ref{Z}).  We next group together factors with the same winding
number.  For given $l$, the right-hand side of Eq.\ (\ref{Z}) has $l$
factors, which can be permuted in $l!$ possible ways.  If a
permutation contains $r_j$ identical factors with winding number $j$, we
have to divide by $r_j!$ because they are indistinguishable. Consequently,
\begin{eqnarray} \label{Zcycl}
Z \hspace{-.25cm} &=& \hspace{-.25cm} \sum_{l=1}^\infty
\frac{1}{l!}\sum_{N=1}^\infty \sum_{\begin{array}{c} \scriptstyle
r_1,\dots,r_N=0 \\ \scriptstyle \sum_j r_j=l \\ \scriptstyle \sum_j j r_j =
N \end{array}}^\infty \frac{l!}{\prod_j r_j!}  \left[V \eta^{1+1}
G_1(0)/1\right]^{r_1} \cdots \left[V \eta^{N+1} G_N(0)/N\right]^{r_N}
\nonumber \\ \hspace{-.25cm} &=& \hspace{-.25cm} \sum_{N=1}^\infty
\frac{1}{N!}  \sum_{\begin{array}{c} \scriptstyle r_1,\dots,r_N=0 \\
\scriptstyle \sum_j jr_j=N
\end{array}}^\infty \frac{N!}{\prod_j j^{r_j}r_j!}  \left[V
\eta^{1+1}G_1(0)\right]^{r_1} \cdots \left[V\eta^{N+1}
G_N(0)\right]^{r_N}, 
\end{eqnarray} 
where in the last equation we carried out the sum over $l$ thereby
losing the constraint $\sum_j r_j=l$, and inserted unity $1=N!/N!$.
The connection with Feynman's path-integral representation of $Z$ is
established by noting that a propagator with winding number $j>1$ can be
represented as a product of $j$ propagators each of unit winding number:
\begin{eqnarray} \label{inter}
\lefteqn{V G_j(0) = \int \dd^3 x_1 \int \dd^3 x_2 \cdots \int
\dd^3 x_j \; G_1[(0,{\bf x}_1),(\hbar \beta,{\bf x}_2)]} \\ &&
\hspace{1.5cm} \times G_1[(\hbar \beta,{\bf x}_2),(2\hbar \beta,{\bf x}_3)]
\cdots G_1[((j-1)\hbar \beta,{\bf x}_j),(j \hbar \beta,{\bf x}_1)],
\nonumber
\end{eqnarray} 
where we inserted $j-1$ intermediate points ${\bf x}_2 \cdots {\bf
x}_j$, and integrated over them so as to allow for all possible values
of the intermediate positions.  The integral $\int \dd^3 x_1$ yields the
volume factor at the left-hand side of Eq.\ (\ref{inter}).  Owing to the
periodicity in Euclidean time $\tau$ with period $\hbar \beta$, Eq.\
(\ref{inter}) may be equivalently written as:
\begin{eqnarray} \label{interred}
\lefteqn{V G_j(0) = \int \dd^3 x_1 \int \dd^3 x_2 \cdots \int
\dd^3 x_j \; G_1[(0,{\bf x}_1),(\hbar \beta,{\bf x}_2)]} \\ &&
\hspace{1.5cm} \times G_1[(0,{\bf x}_2),(\hbar \beta,{\bf x}_3)] \cdots
G_1[(0,{\bf x}_j),(\hbar \beta,{\bf x}_1)]. \nonumber
\end{eqnarray}
Physically, as was first observed by Montroll and Ward \cite{MoWa}, a single
loop with winding number $j$ is reinterpreted as describing $j$ particles
which after a time $\hbar \beta$ are being cyclicly permuted: $1 \rightarrow
2, \, 2 \rightarrow 3, \, \cdots, \, j \rightarrow 1$ (see Fig.\
\ref{fig:many}).

Amusingly, this picture is reminiscent of Wheeler's attempt to explain why
all electrons have the same mass and charge by supposing that their
worldlines form a single huge knot \cite{Schweber}.  When the knot is cut by
a hyperplane of fixed time, many worldlines corresponding to many electrons
would appear.  The flaw in Wheeler's argument is---as was pointed out to him
by Feynman---that to each worldline going forward in time there should be
one going backward in time, corresponding to a positron.  This would imply
an equal number of electrons and positrons, which is not realized in nature.
Our picture is not nullified by this critique because, owing to the
periodicity in $\tau$, the worldlines discussed here always move forward in
time.

\begin{figure}
\begin{center}
\epsfxsize=5.cm
\mbox{\epsfbox{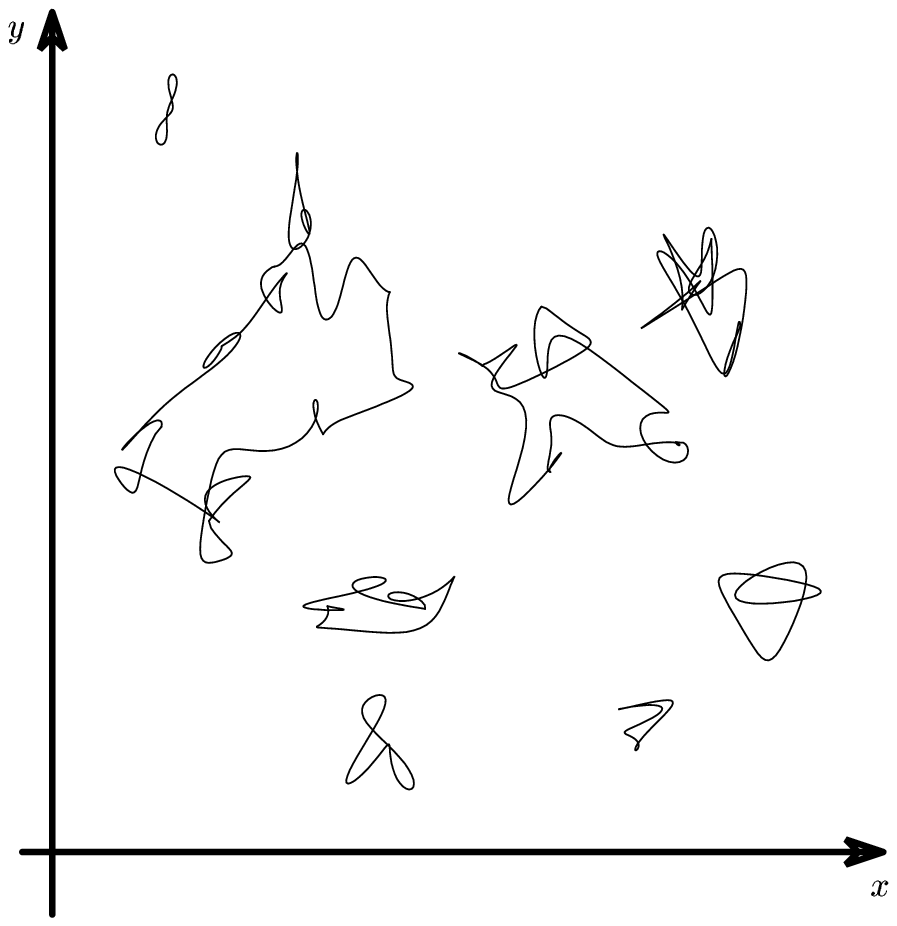}}
\hspace{3em}
\epsfxsize=5.cm
\mbox{\epsfbox{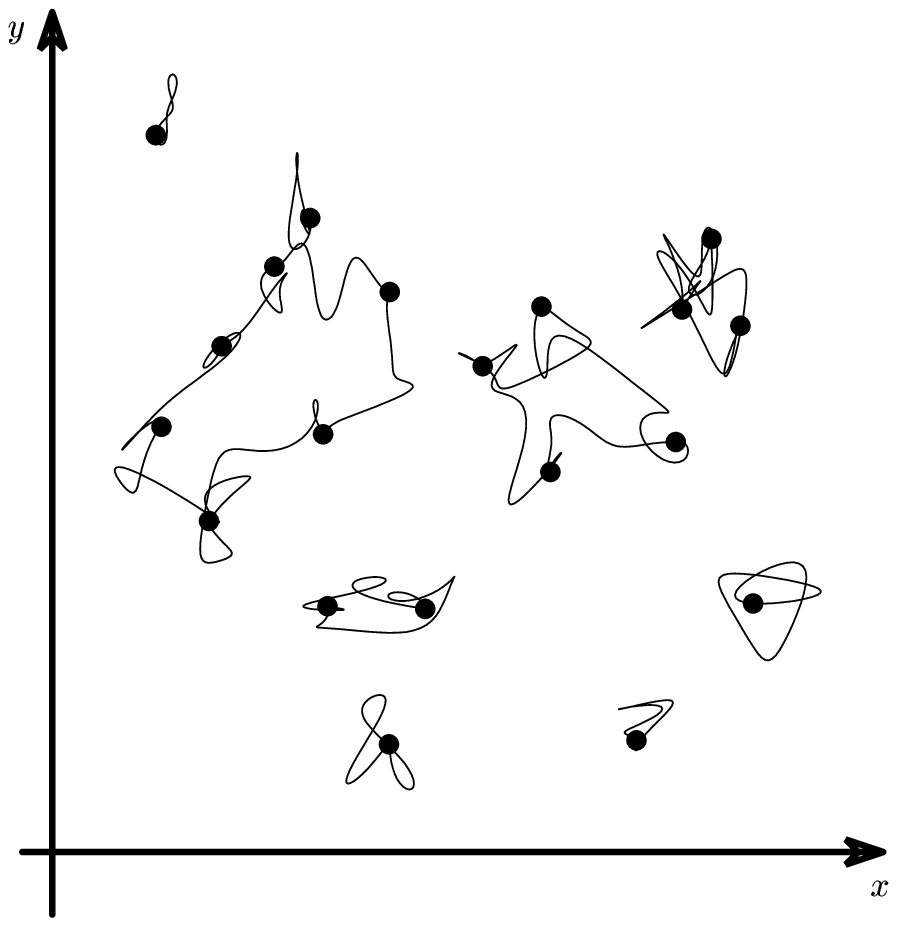}}
\end{center}
\figcap{Connection between the string (left panel) and cycle representation
(right panel) of the partition function. \label{fig:loop-cycle}}
\end{figure}

We recognize in Eq.\ (\ref{Zcycl}) apart from the identity permutation
$G_1(0)$, an interchange of two particles $G_2(0)$, a three-particle cyclic
permutation $G_3(0)$, and so on.  Since $\sum_j j r_j=N$, the integer $N$,
which was introduced in Eq.\ (\ref{extra}), denotes the total number of
particles present in the cycles.  Figure \ref{fig:loop-cycle} gives a
two-dimensional illustration of how the loop and cycle interpretation are
connected.  In the left panel, traces of eight loops in the $xy$-plane are
depicted, each of which is parameterized by the Euclidean time $\tau$.  A
closer inspection reveals that, for example, the longest loop has winding
number $w=7$.  In the right panel, a filled circle is drawn whenever a time
$\hbar \beta$ elapsed.  They mark the starting point of a random walk
executed by one particle during the time $0 \leq \tau \leq \hbar \beta$ and
the end point of one executed by another particle.  In the longest loop, 7
particles are cyclicly permuted in this way.

Now, any permutation of $N$ elements can be factored into cycles of length
$r_j$ $(j=1,2, \cdots, N)$.  The number $M(r_1, r_2, \cdots, r_N)$ of such
permutations is given by \cite{statmech}
\begin{equation} 
M(r_1, r_2, \cdots, r_N) = \frac{N!}{\prod_j j^{r_j}r_j!},
\end{equation} 
which is precisely the combination appearing in (\ref{Zcycl}).  In this way,
we arrive at the well-known path-integral representation of the partition
function due to Feynman \cite{lambda,statmech}:
\begin{eqnarray} 
\lefteqn{Z = \sum_{N=1}^\infty \frac{1}{N!}  \sum_{\sigma\in S(N)} {\rm
sgn}^{\eta'}(\sigma) \int \dd^3 x_1 \cdots \dd^3 x_N G_1[(0,{\bf
x}_{\sigma(1)}),(\hbar \beta,{\bf x}_{\sigma(2)})]} \\ &&
\hspace{2.0cm} \times G_1[(0,{\bf x}_{\sigma(2)}),(\hbar \beta,{\bf
x}_{\sigma(3)})] \cdots G_1[(0,{\bf x}_{\sigma(N)}),(\hbar \beta, {\bf
x}_{\sigma(1)})], \nonumber
\end{eqnarray} 
where the sum $\sigma$ is over all permutations $S(N)$ of $N$ particles, and
\begin{equation} 
\eta' = \frac{1}{2}(1-\eta) = \left\{ \begin{array}{ll} 0 & \mbox{for
bosons} \\ 1 & \mbox{for fermions}.  \end{array} \right.
\end{equation} 
(A similar representation of $Z$, but this time not involving path
integrals, was obtained earlier by Matsubara \cite{Matsubara}.)  We thus
have established the relation between the string picture and Feynman's,
where BEC is connected with the appearance of large rings involving the
cyclic permutation of many particles.

In conclusion, we have described a nonrelativistic Bose gas as a path
integral over spacetime loops.  The loops are characterized by their string
tension and the number of times they wind around the imaginary time axis.
The winding number of a given loop corresponds to the number of cyclicly
permuted particles in a ring in Feynman's theory.  The string tension, which
is determined by the chemical potential, exhibits critical behavior at
$T_{\rm c}$.  Owing to the vanishing of the string tension, the loops
proliferate, meaning that for an infinite system, loops with arbitrary large
winding numbers appear \cite{Stone}.  Particles contained in long loops were
shown to be condensed in the ground state.  \\

\begin{center}
{\bf Acknowledgment} 
\end{center}
We thank N. Antunes, L. Bettencourt, H. Kleinert, A. Nguyen, and
A. Sudb{\o} for discussions and correspondence.  This work is performed
as part of a scientific network supported by the European Science
Foundation, an association of 62 European national funding agencies (see
network's URL, {\tt http://www.physik.fu-berlin.de/ $\sim$defect}).


\begin{thebibliography}{99}
\bibitem{Boulder} H. M. Anderson, J. R. Ensher, M. R. Matthews,
C. E. Wieman, and E. A. Cornell, Science {\bf 269}, 198 (1995).
\bibitem{MITBEC} K. B. Davis, M.-O. Mewes, M. R. Andrews, N. J. van Druten,
D. S. Durfee, D. M. Kurn, and W. Ketterle, Phys. Rev. Lett. {\bf 75}, 3969
(1995). 
\bibitem{Huang} K. Huang, {\it Statistical Mechanics} (Wiley, New York,
1987).
\bibitem{Bogoliubov} N. N. Bogoliubov, J. Phys. USSR {\bf 11}, 23 (1947).
\bibitem{Brown} L. S. Brown.  {\it Quantum Field Theory} (Cambridge
University Press, Cambridge, 1992).
\bibitem{lambda} R. P. Feynman, Phys. Rev. {\bf 90}, 1116 (1953); {\it
ibid} {\bf 91}, 1291 (1953).
\bibitem{Feynman48} R. P. Feynman, Rev. Mod. Phys. {\bf 20}, 367 (1948).
\bibitem{statmech} R. P. Feynman, {\it Statistical Mechanics}
(Benjamin, Reading, 1972).
\bibitem{Stone} M. Stone, Int. J. Mod. Phys. B {\bf 4}, 1465 (1990).
\bibitem{ChanWol} D. Chandler and P. G. Wolynes, J. Chem. Phys. {\bf 74},
4078 (1981).
\bibitem{CePo} D. M. Ceperley and E. L . Pollock, Phys. Rev. Lett. {\bf
56}, 351 (1986).
\bibitem{Ceperley} D. M. Ceperley, Rev. Mod. Phys. {\bf 67}, 279 (1995).
\bibitem{MoWa} E. W. Montroll and J. C. Ward, Phys. Fluids {\bf 1}, 55
(1958).
\bibitem{ABH} N. D. Antunes, L. M. A. Bettencourt, and M. Hindmarsh,
Phys. Rev. Lett. {\bf 80}, 908 (1998); N. D. Antunes and
L. M. A. Bettencourt, Phys. Rev. Lett. {\bf 81}, 3083 (1998)
\bibitem{NgSu} A. K. Nguyen and A. Sudb{\o}, {\it Topological phase
fluctuations, amplitude fluctuations, and criticality in extreme type-II
superconductors}, preprint (1999).
\bibitem{GFCM} H. Kleinert, {\it Gauge Fields in Condensed Matter} (World 
Scientific, Singapore, 1989) Vol.~1.
\bibitem{Rivers} R. J. Rivers, {\it Path Integrals in Quantum Field
Theory} (Cambridge University Press, Cambridge, 1987).
\bibitem{proptime} J. Schwinger, Phys. Rev. {\bf 82}, 664 (1951).
\bibitem{Copetal} E. Copeland, D. Haws, S. Holbraad, and R. Rivers, in:
{\it The Formation and Evolution of Cosmic Strings}, edited by
G. W. Gibbons, S. W. Hawking, and T. Vachaspati (Cambridge University
Press, Cambridge, 1990).
\bibitem{Schweber} S. S. Schweber, Rev. Mod. Phys. {\bf 58}, 449 (1986).
\bibitem{Matsubara} T. Matsubara, Prog. Theor. Phys. {\bf 6}, 714 (1951).
\end{thebibliography}
\end{document}